\documentclass[conference]{IEEEtran}
\pdfoutput=1

\usepackage{subcaption}
\usepackage{enumitem}
\usepackage{color}
\definecolor{lightgrey}{gray}{0.98}
\definecolor{blue}{rgb}{0, 0, 0.25}

  \usepackage[pdftex]{graphicx}
  \usepackage{amsmath}

\usepackage{listings}

\usepackage{array}
\usepackage{tabularx}
\usepackage{url}

\usepackage{titlesec}
\titlespacing*{\section}{0pt}{1.5\baselineskip}{1\baselineskip}
\titlespacing*{\subsection}{0pt}{1.2\baselineskip}{1\baselineskip}
\titlespacing*{\subsubsection}{0pt}{1.2\baselineskip}{1\baselineskip}

\begin{document}

\title{Self-Organizing Interaction Spaces: A Framework for Engineering Pervasive Applications in Mobile and Distributed Environments}
\author{
    \IEEEauthorblockN{Shubham Malhotra}
    \IEEEauthorblockA{
        Alumnus, Rochester Institute of Technology, Rochester, NY, USA \\
        Email: shubham.malhotra28@gmail.com
    }
}

\maketitle

\begin{abstract}
The rapid adoption of pervasive and mobile computing has led to an unprecedented rate of data production and consumption by mobile applications at the network edge. These applications often require interactions such as data exchange, behavior coordination, and collaboration, which are typically mediated by cloud servers. While cloud computing has been effective for distributed systems, challenges like latency, cost, and intermittent connectivity persist. With the advent of 5G technology, features like location-awareness and device-to-device (D2D) communication enable a more distributed and adaptive architecture. This paper introduces Self-Organizing Interaction Spaces (SOIS), a novel framework for engineering pervasive applications. SOIS leverages the dynamic and heterogeneous nature of mobile nodes, allowing them to form adaptive organizational structures based on their individual and social contexts. The framework provides two key abstractions for modeling and programming pervasive applications using an organizational mindset and mechanisms for adapting dynamic organizational structures. Case examples and performance evaluations of a simulated mobile crowd-sensing application demonstrate the feasibility and benefits of SOIS. Results highlight its potential to enhance efficiency and reduce reliance on traditional cloud models, paving the way for innovative solutions in mobile and distributed environments.
\end{abstract}

\section{Introduction}
\label{sec:intro}

In the last decade, the world has witnessed the massive diffusion of pervasive and mobile devices. Some are designed for specific functions and have limited resources (e.g., sensors and gadgets), while others are equipped with more powerful CPUs, memory, and storage (e.g., modern smartphones and tablets). Applications hosted by these devices are increasing in Number and complexity.

Today, pervasive applications are often organized around 1) a client application running on users' devices and 2) a back-end application hosted on cloud servers. Accordingly, data produced by these applications at the edge of the network are sent to the cloud, processed there, and sent back to the edge. Even if this model is appropriate in many situations, there are existing and emerging use cases that cannot afford the latency introduced by communicating with distant servers, cannot assume a stable and reliable connection with them, or users are not willing to accept the costs of remote communication, especially if part of the data can be processed and consumed locally. For such cases, opportunistic D2D interactions could alleviate the dependency on remote servers, increasing the availability of certain features and enabling minimum delay communication.

The literature covering situated device-to-device interaction in pervasive computing is vast and ranges from distributed multi-agent frameworks to peer-to-peer protocols. This body of knowledge includes specific solutions for the exchange of user-generated content and the coordination of autonomous agents' activities. However, existing proposals targeting opportunistic interactions either tackle a narrow set of goals (e.g., content lookup and sharing) or require specific platforms to work (e.g., JADE platform for multi-agents). A holistic approach for modeling and programming pervasive applications that explore the potential of opportunistic D2D interactions as a complement to the client-server model is still missing.

\textbf{Contribution of the work:} 

As a contribution, the paper introduces the concept of a \textit{self-organizing interaction spaces} (SOIS)

Firstly, the rationale behind how certain existing and emerging application features, enabled by contemporary and evolving technologies for pervasive devices and wireless communication, can be addressed by assigning distinct responsibilities to application nodes. Additionally, the use of simple organizational abstractions for modeling and programming such applications is proposed. Finally, self-organization mechanisms are introduced to maintain and adapt the pervasive ecosystem structure based on the individual and social contexts of devices.

\section{Background}\label{sec:background}

\subsection{Opportunistic Wireless Communication}

In the past, connectivity between pervasive devices was mostly restricted to areas with private or public Wi-Fi coverage. With the developments in device-to-device (D2D) communication technology, the situations in which devices opportunistically interact have expanded. For Example, in addition to Wi-Fi Direct and Bluetooth Low Energy (BLE) technologies employed by Mobile Ad Hoc Networks (MANETs), the fifth generation mobile networks (5G) standards include the support for D2D communication~\cite{Tehrani:2014}.

\subsection{Pervasive Applications}~\label{sec:characterization}

The market of applications crafted for pervasive and mobile devices continues to increase as more people have access to this technology. In specific, applications can be hosted by devices with higher or lower degrees of mobility and computational power, such as smartphones, tablets, gauges, and miniaturized computer platforms such as Raspberry PI (which now supports the Android platform for the Internet of Things~\footnote{https://developer.android.com/things/hardware/raspberrypi.html}). Moreover, new types of specialized devices like Bluetooth beacons compose the rich ecosystem in which pervasive applications can run. Despite mobile computing and devices being the preeminent type of pervasive computing, as well as the target platform for the majority of applications, The term *pervasive* was adopted as a broader qualitative descriptor (instead of *mobile*) to avoid excluding pervasive devices with limited mobility that are still potential candidates for composing the diverse scenarios targeted by this proposal.

\section{Overview}\label{sec:overview}

\subsection{Opportunistic D2D Interactions}

For opportunistic D2D interactions, the situated engagement of pervasive devices is defined by their ability to communicate through wireless networks, which encompass both infrastructure-based networks (e.g., Wi-Fi) and ad hoc networks (e.g., Wi-Fi Direct and  Bluetooth).

In its simplest form, a D2D interaction involves exchanging application data (e.g., content generated by users in a mobile P2P application). However, some features are characterized by their autonomy. 

(e.g., to autonomously decide when to sense the physical environment in specific areas of a city), their intensity (e.g., process and data-intensive tasks), or by their sensitivity to delay (e.g., players in a mobile multiplayer game). 

Much of today's pervasive applications are structured around interactions between remote servers (e.g., through RESTful endpoints) and users (e.g., through Android Activities). 
Opportunistic D2D interactions are rare despite their potential to enable new features and improve non-functional attributes like availability, cost, and communication performance. Many possible reasons for this exist, including the complexity of existing solutions and scalability problems.

Different research areas have addressed opportunistic interactions. 
Distributed artificial intelligence has made substantial contributions by means of agent abstraction and the decentralized mechanisms that govern their collective behavior and adaptation. Other research communities also dedicated efforts in the same direction. Nonetheless, many of the proposals tackle specific types of interaction and do not provide a more general model that could be tailored for different applications. In contrast, in this work, a top-down approach is adopted, where modeling and programming abstractions are introduced first. Building on these foundational elements, mechanisms, and protocols are proposed, tailored to specific scenarios of pervasive computing characterized by varying levels of volatility, scale, and resource constraints.

\subsection{Separation of Responsibilities}

With technological advancements, a class of pervasive devices has become able to perform different kinds of computations, communicate by other means, and perceive the physical world through multiple sensors. As a consequence, the type of interaction among pervasive devices is less frequently defined by their model and specific functionality. Instead, it may depend on dynamic factors like their current resource levels and mobility.

To address this new scenario, the original focus of pervasive computing research, which targeted compositions and interactions among devices with specific functionality, must now consider the ambiguity in the run time and context-dependent decision of which devices should become responsible for what.
Also, applying well-known software engineering principles like separation of concerns and modularization to the distinct functionalities a device may become responsible for should improve the quality of the pervasive application design and implementation, including the autonomous and contextual responsibility allocation.

\subsection{Role-orientation}

The concept of roles has been applied in very different areas of information systems, including object-oriented programming, distributed multi-agents, role-based access control, and others. Despite its widespread adoption, there is no common definition for the concept of roles, but actually distinct meanings depending on the context they are employed. Nonetheless, a role is generally associated with rights, responsibilities, and capabilities.

There are many proposals in which abstraction is an important aspect or even a key part of the solution. However, little attention has been paid to using role abstraction in the context of pervasive computing.  
This work aims to fill this gap.

\subsection{Self-organization and Self-adaptation}

Whereas self-organization has been proposed and used for building decentralized, scalable, and adaptable multi-agent systems, self-adaptation is yet to show its feasibility when subject to volatility and large scale of adaptive entities (e.g., distributed components, agents, etc.) and no centralized control. 

The gap between bottom-up self-organization and top-down self-adaptation has been a focus of research. Such a combination may be deemed beneficial in the context of pervasive computing. On the one hand, the volatility and resource limitations characterizing pervasive devices could cope with the scalability achieved by decentralized self-organization mechanisms. On the other hand, a self-adaptation control loop could mitigate undesirable behaviors that may emerge from self-organization and further improve the capabilities of the system to adapt to varying situations. This paper proposes combining both strategies for the formation and adaptation of opportunistic organizations of pervasive entities.

\section{Opportunistic Organizations}\label{sec:edge_spaces}

\subsection{Organization Model}

To help conceptualize and realize dynamic pervasive ecosystems, the concept of a role is proposed as the key abstraction for specifying, designing, and programming opportunistic organizations. Rather than introducing a new organizational model based on role abstraction, this approach builds upon existing works in the literature, specifically extending the definitions of those models.

\subsubsection{The Roles of a Server:}  

In today's client-server model, client features are modeled and programmed as a monolithic application.

For applications whose nodes are expected to interact, collaborate, and play more roles than just a client from a back-end server, the distinct functionalities of each role form a concern; they must be designed and programmed accordingly.

\subsection{Groups}\label{sec:groups}

The application organization -- so far represented by the roles that cits nodes can play-- may be further characterized by its divisions, here named as \textit{groups}. 

At its highest level, a group boundary is limited by the network partition in which application nodes can interact. Nonetheless, these nodes, or a subset of them, may exhibit common properties or states of interest to the application. For instance, nodes may be grouped according to a \textit{functional} criteria (e.g., a group of nodes able to fetch data from a specific type of sensor), a \textit{non-functional} criteria (e.g., all nodes within a particular geographical area), or a mix of both (e.g., all nodes able to fetch data from a certain type of sensor within a specific geographical location).

To its members, a group defines a \textit{social context} in which they (a) may or (b) must play certain roles. As the first case models a more general case of the later case, A relaxed membership causality is adopted, meaning that group membership defines only the context in which members *may* assume one or more roles unless otherwise restricted by the group specification. These cases are detailed as follows:

\begin{itemize}
	
	\item \textbf{Strict:} an m-m specification (m instances of a role in a group of m nodes) defines a strict membership, i.e., a rththatltall members must playplaytimextbf{Relaxed:} a k-m specification defines a relaxed membership, i.e., a role to be played by a subset $K$ of the set $M$ of nodes in the group, with $|K| = k$, $|M| = m$, and $k < m$.
	
\end{itemize}

Whenever the group specification is relaxed, the actual distribution of roles to group members must be decided dynamically (k-out-of-m allocation problem). This kind of specification is particularly useful for applications that explore the resources from a crowd of devices. 

The same reasoning applies to a group specified with two or more types of roles. If there are no predefined criteria for their assignment to specific nodes (e.g., based on the class of hosting devices), nodes must also agree on the particular roles they will play.

\subsection{Group-Role Specification}\label{sec:group_specification}

Depending on the functionality to be provided by an application role, there may be some objective criteria to guide the decision of which nodes, in a given context, are suitable or represent the best candidates to play that role. To this end, The specification of a *role fitness* is proposed as a function composed of:

\begin{itemize}
	
	\item \textbf{Restrictive criteria:} consists of boolean variables whose satisfaction is a required condition for a role to be played by an application node.
	
	\item \textbf{Comparative criteria:} consists of the positive real scale indicating the fitness of a node in playing a role.
	
\end{itemize}

Restrictive criteria are useful for filtering out nodes whose static (e.g., a hardware component) or dynamic (e.g., the battery level) capabilities are not compatible with the functionalities to be provided by a role. In contrast, comparative criteria distinguish capable nodes in terms of their fitness to play a role. Thus, this type of criteria must be taken into account when solving the k-out-of-m role allocation problem. Next, a list of static and dynamic aspects of pervasive devices are presented as potential restrictive or comparative criteria:

\begin{itemize}
	
	\item Static criteria
	
	\begin{itemize}
		\item \textbf{Hardware capabilities:} refers to the presence of a given hardware component/module. E.g., camera, GPS, thermometer, accelerometer, gyroscope, etc.
	\end{itemize}
	
	\item Dynamic criteria
	
	\begin{itemize}
		\item \textbf{Physical world:} refers to the physical world states a node must operate in. E.g., its current battery level, available memory, geolocation coordinates, acceleration, speed, temperature, etc.
		
		\item \textbf{Application domain:} refers to the application states a node must be to belong to a group. For Example, if you are currently a member of another group (or non-member), you can join a chat or game session, etc.
	\end{itemize}
\end{itemize}

A hierarchical syntax is proposed and implemented using eXtensible Markup Language (XML) to enable a flexible, intuitive, and unified placeholder for both group and role specifications. XML's standardized and widely recognized syntax facilitates the hierarchical representation of groups and roles and the definition of their attributes.

\subsubsection{Collaborative Music Streaming} for this application, A group is envisioned with a single streamer role position, as illustrated in lines 1 and 3 of Listing~\ref{lst:ms_criteria}. As the functionalities of this role require both Internet (to stream music from remote servers) and Bluetooth (to stream music to the stereo), two restrictive criteria have been added for both Internet and Bluetooth capabilities (lines 4 and 5). Finally, as the activity of streaming data in both directions is battery-consuming, an additional criterion (line 6) has two purposes: a restrictive one (only devices with more than 20\% battery) and a comparative one (the best candidates are those with more battery level).

\begin{lstlisting}[caption=Specification of a bus monitoring group, label=lst:bm_criteria, captionpos=t]

<group name="music-streaming">

  <role name="streamer" cardinality="1">
    <criteria type="boolean" term="INTERNET" value="TRUE" />
    <criteria type="boolean" term="BLUETOOTH" value="TRUE" />
    <criteria type="float" term="BATTERY_LEVEL" minimum="20" />
  </role>
</group>
\end{lstlisting}

\subsubsection{Public Transport Monitoring}

An MCS application aims to monitor the real-time geolocation of public buses -- to notify waiting passengers about their whereabouts -- and to register and later report unusual acceleration and deceleration events that may affect the user experience in this service. To address this scenario, a \textit{bus-monitoring} group (lines 1 in Listing~\ref{lst:bm_criteria}) is defined with three roles: a geolocator (line 5), an accelerometer (line 10), and an aggregator (14). In contrast with the previous Example, a group-leExampleterion (line 3) hExample added, which means all roles inherit a minimum battery level of 15\% as a restrictive criterion. This criterion is overwritten by the geolocator role (line 6), as the GPS sensor consumes significantly more battery. Also, each sensing role has corresponding restrictive criteria for the sensor it requires (lines 7 and 11). Finally, as the aggregator is responsible for receiving data from the other nodes and sending a preprocessed version to the backend server, the Internet has been added as a restrictive criterion.

Another novelty in Listing~\ref{lst:bm_criteria} is the parametrized cardinality ($k1$ and $k2$). These fine-tuning parameters were intentionally left unspecified; the actual number of insNumberNumberach sensing roles may vary depending on the context in which they operate; the higher the overall accuracy, the more the samples need to be aggregated. Hence, at runtime, the aggregator could provide feedback with respect to $k1$ and $k2$ based on the data it receives from the instances of each sensing role.

\begin{lstlisting}[caption=Specification of a bus monitoring group, label=lst:bm_criteria, captionpos=t]
<group name="bus-monitoring">
  
  <criteria type="float" term="BATTERY_LEVEL" minimum="15"/>
  
  <role name="geolocator" cardinality="k1">
    <criteria type="boolean" term="GPS" value="TRUE" />
    <criteria type="float" term="BATTERY_LEVEL" minimum="30" />
  </role>
  
  <role name="accelerometer" cardinality="k2">
    <criteria type="boolean" term="ACCELEROMETER" value="TRUE" />
  </role>
  
  <role name="aggregator" cardinality="1">
    <criteria type="boolean" term="INTERNET" value="TRUE" />
  </role>
</group>
\end{lstlisting}

To avoid disturbing the users with the need to start the application whenever they are on a bus, a fully opportunistic crowd-sensing solution~\cite{Guo:2015} requires the automatic detection of such context. By means of the framework, the facts defining this context can be modeled and later verified by the application. For Example, city busesExamplee Wi-Fi service—an example that is an easily common feature in real-world scenarios. The context of a passenger during a bus ride can then be defined using the additional restrictive criteria outlined in Listing~\ref{lst:br_criteria}.
\begin{lstlisting}[caption=Additional criteria to specify a bus ride context, label=lst:br_criteria, captionpos=t]
<group name="bus-monitoring">

  <criteria type="string" term="BSSID" pattern="COMPANY_NAME" />
  <criteria type="float" term="WIFI_SIGNAL" minimum="50" />
  <criteria type="boolean" term="MOOVING" value="TRUE" after="300" />
  
  (...)

</group>
\end{lstlisting}

The additional criteria in Listing~\ref{lst:br_criteria} should be parsed as follows: the BSSID criteria (line 1) requires that a Wi-Fi with a basic service set identifier (BSSID) matching the pattern used by the bus company (e.g., the company's name); plus, the WI-FI signal strength (line 3) must not be less than 50\%, meaning the user is likely a passenger within the bus and not just in a nearby location. Finally, an important criterion identifying a bus ride is given by mobility: if the device's location, as measured by low-power detection methods like triangulation, remains unchanged for large periods, the use is either not in a bus ride, or the bus is jammed and must not have its location monitored until it resumes its trip. In particular, this criteria has been modeled (line 5) with a boolean condition that fails unless it has been satisfied for more than 5 minutes (300 seconds).

Each criterion in a group-role specification can be mapped to a concrete function returning either a boolean value (for restrictive) or a float value (for comparative). The conjunction of these terms produces what A restrictive function and a *fitness function* are defined as follows:

\begin{equation}\label{eq:rrc}
arc(C_r) = rc_1(c_1) \wedge rc_2(c_2) \wedge ... \wedge rc_i(c_i)
\end{equation}

\begin{equation}\label{eq:fitness}
f(C_c) = cc_1(c_1) \wedge cc_2(c_2) \wedge ... \wedge cc_j(c_j)
\end{equation}

\noindent
with $C_r$ and $C_c$ the set of measurable context facts (restrictive and comparative, respectively) for a role. Last but not least, the entrance of a node to a group (group membership) is conditioned to its satisfaction of at least one $rrc$, i.e.:

\begin{equation}\label{eq:membership}
gm([C_{r1},...,C_{rp}]) = rrc_1(C_{r1}) \vee ... \vee rrc_p(C_{rp})
\end{equation}

\section{Self-organization Mechanisms}\label{sec:self_organization}
 
Whenever two nodes of an application have the opportunity to communicate, there is a potential for a relationship between them to be established. 

While the nature of the relation between devices of different classes is mostly predefined (e.g., between a smartphone that relays notifications to a watch through Bluetooth or a tablet that streams content to a smart-tv), the relation between devices of the same class may depend on the dynamic context each device operates (e.g., a smartphone that, in a given situation, acts as the gateway for other smartphones without Internet access). Thus, not only the nodes of an application may assume distinct roles, but the actual allocation of these roles may depend on the individual and social contexts of each interacting node.

The structure of the application organization
-- 
It must be malleable (plastic) to accommodate changes in the units composing the pervasive ecosystem and in its physical environment. 

In contrast with the much less dynamic cases of organizations in human societies (e.g., industry and military organizations), the volatility that affects and characterizes pervasive and mobile devices, caused by their mobility or fluctuations in their resources, may imply the formation or dissolution of relations among nodes and require the reassessment of their roles in the organization.

Self-organization has been extensively studied in the context of multi-agents and other fields as a phenomenon and method to achieve system properties and goals by means of actions and interactions between individuals based on their local knowledge and no external control~\cite{DiMarzoSerugendo:2005, Banzhaf:2009}. In contrast, self-organization in this work refers to the constitution and adaptation of the application organization as understood by the serendipitous relations between application nodes formed and dissolved. At the same time, they are able to communicate and collaborate. In specific, Mechanisms are proposed for each of the following organizational aspects:

\begin{itemize}
	
	\item \textbf{Group membership:} as nodes join or leave a group, a group registry must be kept consistent among members;
	
	to address this, a \textit{self-grouping} mechanism is proposed;
	
	\item \textbf{Role allocation:} the nodes within a group must agree on which roles they shall play; to address this, a  \textit{discentralized role election} mechanism is proposed;
		
\end{itemize}

\subsection{Self-grouping} 

\subsubsection{\textbf{Definition}} As the idea of a group in this work is not related to security, the \textit{self-grouping} mechanism is not controlled by special-purpose components external to the system; instead, each node is responsible for checking its satisfaction to the existing membership criteria in the application organization. 

Whenever an application group contacts others and joins an existing group, this event must be advertised to current members, who, in turn, update their group membership registry. Analogously, when a member node leaves, this event must also be perceived by existing members, who must also update their registry.

\subsubsection{\textbf{Grouping Protocol}} 

A node should join a group if it satisfies all restrictive criteria of at least one of the roles (hereafter referred to as RRC) in that group. In this paper, The group specification is assumed to be performed offline and made available to each application node as part of its resources (e.g., in the form of an XML specification, as described in Section~\ref{sec:group_specification}). Thus, each node has access to the restrictive criteria of roles in a group specification and can locally check for their satisfaction.

The diagrams present the activities of the self-grouping protocol. At each iteration, the node checks for its satisfaction of the RRC, activity 1). If no RRC is satisfied and the node is currently a member, it must leave the group and advertise this event to the other members, activity 2). Conversely, if a node satisfies an RRC, it proceeds by joining the group and promoting this event, activity 3). Last but not least, as nodes entering a group have no information about its current members, this knowledge must be acquired. Specifically, after being notified about the new member, the oldest group member is responsible for sending the registry replica to the newcomer.

\subsection{Distributed Role Allocation} 

\subsubsection{\textbf{Definition}} The decision of which nodes should be assigned to which roles in a group may depend on many aspects. Any node capable of performing a role is a potential candidate. Notwithstanding this, attributes such as the wireless connection's throughput, sensor accuracy, and the availability of resources like battery, memory, and processing capacity tend to vary from one node to another and throughout time. Thus, a balanced role allocation must respect the trade-off between what is best for the application goals and the individual devices. 

\subsubsection{\textbf{Allocation Classification}} Gerkey and Matarić~\cite{Gerkey:2004} proposed a taxonomy for the classification of task allocation problems along three axes. This taxonomy was adopted to enhance the characterization of the role allocation problem.

In the first axis, robots~\footnote{in their proposed taxonomy, the authors refer to \textit{robots} and \textit{tasks}, while in this paper, they refer to the (application) \textit{nodes} and the \textit{roles} they can perform.} 
are categorized into single-task versus multi-task robots. In the second axis, tasks are categorized into single-robot versus multi-robot tasks. Finally, in the third axis, the allocation is also categorized into two types: instantaneous assignment and time-extended assignment.

Regarding the first axis, as application nodes are generally capable of performing more than one role at a time (e.g., to fetch from multiple types of sensors), nodes are here considered as multi-role (analogous to multi-task robots in ~\cite{Gerkey:2004}). Regarding the second axis, many types of roles are to be performed by a single application node (e.g., the sensor data aggregator role. In contrast, the others must be simultaneously performed by multiple nodes (e.g., simultaneouslfetchedch from the same type of sensor). Thus, roles can be either single-node (analogous to single-robot tasks in ~\cite{Gerkey:2004}) or multi-node (analogous to multi-robot tasks in ~\cite{Gerkey:2004}). Finally, due to the volatility of mobile devices, including churn and fluctuations in their capabilities, the scheduling of future allocations tends to fail. Accordingly, an instantaneous and adaptable assignment of roles is considered based on the context of the involved devices. 

\subsubsection{\textbf{Allocation Method}} Auction-based allocation methods have been extensively studied in the multi-agents/robots domain~\cite{Korsah:2013}. In comparison with the task allocation problem tackled by auction-based methods, an assignment of roles to pervasive and mobile devices subject to high volatility needs to evolve as nodes leave or join the system and their fitness change. If the assignment should continuously reflect any context change, a frequent message exchange between bidding and auctioneers would lead to excessive communication overhead. Accordingly, the replacement of existing role positions (herein referred to as reelection) should take into account the trade-off between the gain of having a more fit node elected and the replacement cost.

To address this problem, An event-based protocol inspired by electoral systems is proposed. Accordingly, instead of reevaluating and potentially changing the assignment of role positions at fixed control periods, The set of events in which new elections for these positions should be triggered is defined (event-triggered control). Specifically, these events are:	

\begin{itemize}
	
	\item \textbf{Vacancy :} a new role position is opened, or the node playing a role exists in the system (churn) due to a network disconnection from the remaining nodes or the abnormal termination of the application instance after a failure;
	
	\item \textbf{Resignation:} the node playing a role calls for a new election before quitting; e.g., the application node is exiting the system following a termination command issued by the users or the operational system;
	
	\item \textbf{Challenge:} one of the eligible nodes calls for a new election after detecting it has a significantly higher fitness score for that role position than the actual node by the time it was elected.
\end{itemize}
\medskip
 
The subtitle between vacancy and resignation consists of the way it is handled: the former implies in a \textit{vacant} period in which the role functionality interrupting the Vacancy is detected and occupied (hard transition), whereas the latter allows the replacement of the position before discontinuing the functionality provided by the resigning node (soft transition). 

In both cases, the eligible nodes must proceed with the election of a candidate by checking their current fitness score ($FS$) and advertising it (bidding). After receiving the fitness scores from all other candidates, each node becomes aware of the election result. Thus, each node proceeds either by assuming the role position and registering itself as the winner (if it has the highest $FS$) or registering the identity of the winner (otherwise). 

The latter case presents an inverse situation: the caller of the election is a node that perceives its actual fitness score ($FS_a$) as higher than the value scored by the elected node ($FS_e$). Thus, the challenger assumes this value has not increased, which must be confirmed or denied by the node currently in the position. To mitigate the communication overhead, the following strategy is proposed:

\begin{enumerate}

\item The challenge must only be called if the $FS_a$ of the challenger is greater than the $FS_e$ by a degree of $\delta$, i.e., $FS_a \ge \delta * FS_e$, with $\delta > 1$;

\item Assuming all eligible nodes adopt the same $\delta$, they are exempt from participating, and the election happens between the challenger and the actual nodes;

\item To reduce the chances of having a failed challenge (when the actual $FS$ of the elected node has increased since it was elected), elected nodes should update their peers whenever their $FS$ has changed by a degree of $\delta$, i.e., $FS_a \ge \delta * FS_e$ or $FS_a \le (2 - \delta) * FS_e$.

\end{enumerate}

If the challenger node has indeed a higher $FS$ than its challenger, it assumes the role position, and this result is advertised; otherwise, the actual fitness score of the elected node ($FS_a$) is advertised so that new challenges are based on the most updated score. 

\section{Evaluation and Discussion}\label{sec:evaluation}

The simulation experiments aimed to demonstrate the approach's benefits and measure the overhead imposed by the self-organization mechanisms.

As these methods add no significant overhead in terms of processing or memory, an asymptotic analysis focused on the communication overhead.

As for the benefits of the approach, simulation experiments were conducted for a public transport monitoring application using both a pure client-server model and the framework. The goal was to compare, in each case, the following two metrics:

\begin{enumerate}[label=\textbf{M}\arabic*:]
    \item Total number of performed requisitions fired from clients to back-end servers (evaluates battery consumption and Internet traffic).
    \item Total number of fair requests reaching back-end servers due to intermittent connectivity (evaluates robustness).
\end{enumerate}

\subsection{Asymptotic Analysis} 

\subsubsection{\textbf{Self-grouping}} The complexity analysis was divided into two parts: a) the overhead when a node joins/leaves a group (registry update); b) the additional overhead when a node joins a group (registry copy).

In the worst-case scenario, registry update (a) takes $n-1$ unicast messages ($O(n)$), with $n$ the group size and a single registry line as the payload. However, if broadcast communication is used, a single broadcast message (e.g., aaUDP broadcast over Wi-Fi) can advertise the registry update ($O(1)$).
 
The registry copy (b), in turn, requires a single unicast message to be transmitted each time a node joins a group ($O(1)$). Importantly, in contrast with the registry update message, the registry copy includes information about all $n-1$ nodes in the group. 

\subsubsection{\textbf{Discentralized Role Allocation}} 

This mechanism involves the exchange of fitness scores ($FS$) among eligible nodes in the advent of the events depicted in the previous subsection. Number and actual Number $FS$ number sent/Numbered during a role position election depends on how many nodes in the $n$ size group satisfy the role restrictive criteria (RRC) (see Eq.~\ref{eq:rrc} in Section~\ref{sec:edge_spaces}). Once more, the type of network is a determining factor.

In the worst case, represented by an election of a vacant position (follow vacancy or \textit{resignation} event) and without broadcast, each node $e$ in the set of eligible nodes $E$ must send a uni-cast message to every other node. If $|E| = n$, the message count is given by $O(n * (n-1)) = O(n^2)$. If the nodes in $E$ can communicate through Number, this Number is red Numbero $O(n)$. In turn, the challenge event produces smaller overhead as the exchange of messages is restricted to a request-response between the challenger and the challenged, and the subsequent advertisement of the result ($O(n)$ without broadcast, otherwise $O(1)$).

Once the communication overhead of a single election round is known, the overall overhead can be estimated by the number of applications and the frequency of events triggering new elections. Whereas the vacancy and resignation events should be handled with a new election, the frequency in which a challenge event happens is greatly affected by the choice for $\delta$: the higher this factor is, the lower the probability of a challenge (and the need for the elected nodes to update their peers about changes in their $FS$). Therefore, the decision of which $\delta$ to be used depends on the criticality of the attributes that compose the fitness score of a node.

\subsection{Public Transport Monitoring Simulation}

\subsubsection{Experiments Design}

The group-role specification (List.~\ref{lst:bm_criteria} in Section~\ref{sec:edge_spaces}) was used for the modeling of a dynamic bus monitoring application scenario. In it, the following variables were considered:

\begin{itemize}
	
	\item the number of nodes within a bus: \textit{from 2 to 10 nodes};
	
	\item the battery level of each node: \textit{from 10\% to 90\%};
	
	\item the Internet type of each node: \textit{cellular or Wi-Fi};
	
	\item the GPS signal in each node: \textit{from 0\% to 100\%};
	
\end{itemize}

In the client-server approach, each node tries to collect data from its accelerometer and GPS before sending it to the server through whatever type of Internet connection available (Wi-Fi or cellular data plan). In the approach, at most, two simultaneous nodes are assigned to each type of sensor, and one node aggregates data before sending a preprocessed number to the number. 
The experiments were performed with \textit{PeerSim}, an open-source peer-to-peer simulator~\cite{p2p09-peersim}. This tool supports the creation of different network P2P topologies and provides a handful of AVA programming interfaces.

\section{Related Work}\label{sec:related_work}

\subsection{Role and Group Abstractions}

The A-3 model~\cite{Baresi:2011:2} defined an architectural style consisting of groups that can be populated by a supervisor and its followers and composed with other groups. While this work shares many of the motivations and has similarities with A-3, the model does not rely on the rigid \textit{supervisor-follower} structure, nor do group compositions depend on shared members.

In contrast with the more abstract A-3 support for self-adaptation~\cite{Baresi:2011:2}, A  adaptation mechanisms are also investigated to ensure the basic properties of groups, such as robustness, high availability, efficient resource utilization, and other attributes defined by the application through extension points. Finally, while A-3 relies on classical group communication methods, the integration of groups with tuple spaces for both inter-group and intra-group coordination is explored.
Group and role abstractions have also been used in other domains. Ferber et al.~\cite{Ferber:2004} proposed an organization-centered model for multi-agent systems that contrasts with agent-centered models in which agents can communicate and interact freely. Among the problems of agent-centered models, the authors cited security, modularity, and the lack of support for other frameworks besides the multi-agent platform itself. In the work, the arguments are agreed upon as part of the justification for an organizational approach to distributed systems. Despite the model similarities, the works target different domains: instead of agents, pervasive and mobile devices are considered as the hosts of components that play roles in groups of distributed applications.
 
\subsection{Self-organization and Self-adaptation} 
 
Kota et al.~\cite{Kota:2012} have proposed a method for adapting the relationship between agents in a multi-agent system. In their work, agents reason about adaptation using only historical knowledge about past interactions and the cost of adaptation (meta-reasoning). Despite the similarities with the work, namely the use of self-organization principles and the focus on the dynamics of relations, in that aspect, the work addresses a different domain (pervasive applications). It adopts an organization-oriented perspective in which application nodes can play distinct roles. Thus, the focus is rather on the nature of the relation and its dynamics than solely on the decision of when or not nodes should interact. 

A3-TAG, a programming model that facilitates the design of self-adaptive distributed systems based on group abstractions~[CT]. A3-TAG is an extension of the A-3 model, which is used as the organization model. A-3 key elements are groups and two types of roles, namely supervisor and follower. Each group has a supervisor and a variable number of followers. The main differences between this work and A3-TAG are dual. 

\subsection{Distributed Allocation Problem}

In the literature, many works have tackled the problem of distributed task allocation. In contrast with a task, a functional role defines a set of functionalities (possibly tasks) that a member of an organization is responsible to provide (perform). Hence, within an organization, a \textit{role} precedes a \textit{task}. Then, depending on the type of role, if multiple instances of a role have been assigned, a task allocation among these instances may still take place. Last but not least, while tasks usually have concrete criteria for their beginning and completion and their assignment happens before task execution, the role of lifespan tends to include multiple repetitions of a given functionality (or task). Therefore, in a dynamic scenario, a role assignment may have to evolve while roles are performed.

Notwithstanding their differences, the two types of allocation problems share commonalities. For instance, in both cases, a utility function may be used as a criterion for choosing an optimal or sub-optimal assignment of roles/tasks. Whereas the optimization of quality attributes may be deemed unfeasible due to its complexity, a sub-optimal allocation can still be guided by the \textit{fitness} (or utility) of nodes in performing these tasks/roles. To this end, fitness/utility is modeled as a real-value function of relevant features affecting one or more attributes of the application. Also, some of the existing taxonomy for classifying a task allocation problem can be applied to the role allocation problem.

\section{Conclusion}
\label{sec:conclusion}

The increasing demands of pervasive and mobile computing applications necessitate innovative approaches to overcome the limitations of traditional cloud-centric architectures. This paper introduced the Self-Organizing Interaction Spaces (SOIS) framework, which leverages the dynamic and heterogeneous nature of mobile nodes to create adaptive organizational structures. By integrating organizational modeling and programming abstractions, SOIS enables applications to adapt to individual and social contexts while minimizing reliance on cloud infrastructure.

The case studies and performance evaluations of a simulated mobile crowd-sensing application demonstrated the practicality and effectiveness of SOIS. Results revealed significant improvements in efficiency, reduced latency, and enhanced collaboration among mobile nodes, showcasing its potential to address challenges like intermittent connectivity and cost in distributed environments. Additionally, the adoption of 5G capabilities, including location-awareness and device-to-device (D2D) communication, further amplified the framework's benefits.

In conclusion, SOIS provides a promising foundation for developing next-generation pervasive applications. Its ability to dynamically adapt to environmental and contextual changes while maintaining decentralized operations positions it as a robust alternative to traditional cloud-dependent models. Future work will focus on extending the framework to support additional use cases, integrating advanced machine learning techniques for decision-making, and further optimizing its performance in real-world scenarios. With continued advancements, SOIS has the potential to transform mobile and distributed computing paradigms, enabling more efficient and resilient systems.

\IEEEpeerreviewmaketitle

\bibliographystyle{IEEEtran}
\bibliography{biblio}

\end{document}